\documentclass[linenumbers,fleqn,usenatbib]{mnras}

\usepackage{amsmath,amssymb}

\usepackage{xcolor}

\usepackage{graphicx}

\usepackage{hyperref}

\usepackage{multirow}
\definecolor{purple2}{rgb}{0.5, 0.0, 0.5}
\definecolor{red2}{rgb}{0.9, 0.0, 0.1}

\newcommand{\mrm}[1]{\mathrm{#1}}
\newcommand{\nuc}[2]{$\mrm{^{#2}#1}$}

\title[Gamma-Ray Rad Tran Code]{A New Versatile Code for Gamma-Ray Monte-Carlo Radiative Transfer}

\author[Shing Chi Leung]{
Shing-Chi Leung,$^{1}$\thanks{E-mail: leungs@sunypoly.edu}
\\
$^{1}$Department of Mathematics and Physics, SUNY Polytechnic Institute, 
100 Seymour Road, Utica, NY 13502, USA\\
$^{1}$TAPIR, Mailcode 350-17, California Institute of Technology,
Pasadena, CA 91125, USA\\
}

\date{Accepted 14 February 2023. Revised 13 February 2023; in original form 8 May 2022}
\pubyear{2023}

\date{\today}

\begin{document}
\label{firstpage}
\pagerange{\pageref{firstpage}--\pageref{lastpage}}
\maketitle

\begin{abstract}
Ongoing MeV telescopes such as INTEGRAL/SPI and Fermi/GBM, and proposed telescopes including the recently accepted COSI and the e-ASTROGAM and AMEGO missions, provide another window in understanding transients. Their signals contain information about the stellar explosion mechanisms and their corresponding nucleosynthesis of short-lived radioactive isotopes. This raises the need of a radiative transfer code which may efficiently explore different types of astrophysical $\gamma$-ray sources and their dependence on model parameters and input physics. In view of this, we present our new Monte-Carlo Radiative Transfer code in Python. The code synthesizes the $\gamma$-ray spectra and light curves suitable for modeling supernova ejecta, including C+O novae, O+Ne novae, Type Ia and core-collapse supernovae. We test the code extensively for reproducing results consistent with analytic models. We also compare our results with similar models in the literature and discuss how our code depends on selected input physics and setting.

\end{abstract}

\begin{keywords}
transients: novae -- supernovae -- gamma-rays: stars -- nuclear reactions, nucleosynthesis, abundances -- radiative transfer
\end{keywords}


\section{Introduction}
\label{sec:intro}

\subsection{Gamma-Ray Radiative Transfer Code}

$\gamma$-ray spectra provide information about supernovae including the amount of radioactive isotopes and their kinematics \citep{Clayton1974, Isern2021}. Unlike radiative transfer codes for the optical band, there are fewer codes specific for $\gamma$-ray. These codes are in general designed for individual classes of supernovae. 

Monte Carlo schemes are frequently used for hard X-ray and $\gamma$-ray line formation. The particle approach allows direct implementation of the microphysics \citep{Pozdnyakov1983}. For example, the relativistic correction can be applied on the particles in the co-moving frame. Some early works \citep[e.g.,][]{Ambwani1988, The1990} have demonstrated how this approach applies to the $^{56}$Ni and $^{56}$Co decay lines in Type Ia and core-collapse supernovae. 

\citet{Milne2004} compared the performance of different radiative transfer codes designed for $\gamma$-ray transport \citep{Hoeflich1998, Burrows1990, Pinto2001, Kumagai1997, Hungerford2003, Burrows1990}. There exist variations among codes such as (1) how lines are constructed, (2) whether line broadening and density evolution are included, (3) whether relativistic correction is included, (4) which $\gamma$-ray photon interactions are included, and (5) whether the decay of positronium annihilation is included. In the comparison it is shown that variations among these choices do not significantly change the line flux prediction in SNe Ia. The $\gamma$-ray radiative transfer is usually embedded in multi-wavelength radiative transfer codes for modeling core-collapse supernovae \citep{Maeda2006} or Type Ia supernovae \citep{Summa2013}.

\subsection{$\gamma$-ray Astronomy and Sources}

High-energy $\gamma$-ray photons in the MeV range are abundantly produced by a few classes of transient events, including novae, Type Ia supernovae (SNe Ia), core-collapse supernovae (CCSNe) and collapsars. 

Novae are the thermonuclear runaways of C+O or O+Ne white dwarfs (WDs) by accretion from their companion stars, which can be evolved main-sequence stars, He stars or red supergiants \citep[see, e.g., reviews from][]{Starrfield2016, Chomiuk2020}. 
The accretion triggers a thermonuclear runaway (TNR) of the surface H-/He-rich matter \citep[][]{Webbink1987} and the subsequent mass loss \citep{Kato1994}. The ejecta contains matter from the accretion disk and the WD, due to various mixing processes \citep{Goldreich1967, Fujimoto1993, Miles1961, Townsend1958}. The nuclear reactions are mostly $p$-capture and hot-CNO cycle of the C+O-rich and O+Ne-rich matter \citep{Harris1991, Tajitsu2015, Weiss1990} up to the intermediate mass elements (e.g., Ar, Ca) \citep{Jose2001}. Some of the isotopes are radioactive, such as $^{7}$Be, $^{13}$N, $^{18}$F, $^{22}$Na \citep{GomezGomar1998}. Although the isotopic masses are very low ($\sim 10^{-12} - 10^{-7}~M_{\odot}$) \citep{Hernanz1996, Jose2020} and they have a short half-life from minutes to days,
the gamma-rays can escape directly because they are synthesised near the surface. 

SNe Ia are the thermonuclear explosions of C+O WDs in a binary system \citep[see reviews e.g.,][and the reference therein]{Hillebrandt2000, Nomoto2017, Nomoto2018}, where the nuclear runaway is triggered by mass accretion from their evolved binary
or by dynamical ignition during a binary WD merger \citep{Pakmor2010, Pakmor2011}. 
The nuclear burning synthesizes radioactive $^{56}$Ni ($\sim 0.1-0.8~M_{\odot}$) in general \citep{Colgate1969, Chevalier1981, Stritzinger2006, Taubenberger2017}, 
During their decay into $^{56}$Co and then $^{56}$Fe, energetic $\gamma$-ray photons are emitted and escape
when the ejecta becomes optically thin at weeks after the explosion. 
SN Ia has a diverse explosion mechanisms,  e.g., laminar flame \citep{Timmes1992}, turbulent flame \citep{Woosley1997, Reinecke1999, Schmidt2006, Roepke2007def, Leung2020Iax}, double-detonation \citep{Fink2007, Sim2010, Moll2013, Leung2021Sub} and detonation transited from flame \citep[aka the deflagration-detonation transition model, ][]{Khokhlov1991, Golombek2005, Roepke2007, Leung2018}. These models are inspired by the observed diversity in SNe Ia \citep[see e.g.,][]{Leung2021SNIaReview}. 

CCSN is the explosion of a massive star $> 10~M_{\odot}$ powered by neutrino energy deposition after its gravitational collapse \citep{Nomoto1988, Woosley1995, Heger2003, Sukhbold2016}. 
Neutrino energy deposition is an important energy source for its explosion \citep[see recent reviews e.g., ][]{Janka2017} where neutrinos are also responsible for some important processes, $(\nu,p)$-process \citep{Woosley1990} and $r$-process \citep{Woosley1994}.
Most early radiation comes from decay of the isotopes $^{44}$Ti (half life $\tau \sim 60$ years), $^{56}$Ni and $^{56}$Co ($\tau=111.4$ days), $^{60}$Co ($\tau \sim$ 5.71 years). These isotopes are synthesized in the interior of the ejecta. Hence, $\gamma$-ray lines can be seen from months to years after the explosion. 

Collapsar is the explosion of a massive star powered by the center black hole \citep{Woosley1993}. The accretion disk launches jets by magneto-rotational instability \citep{Tsuruta2018}. The jets create high entropy environment for nucleosynthesis and a cone-shape outflow \citep{Tominaga2007, Tominaga2009}. 
After the shock breakout, the opening exposes the central compact object \citep{Zhang2003}, which allows $\gamma$-rays to directly escape.

\subsection{Observations and Motivation}


Nearby novae 
are major candidates for detecting their $\gamma$-rays emitted 
by radioactive isotopes, e.g., the 487 keV line from $^{7}$Be, the 511 keV line from $\beta^+$-decay of short-lived radioactive isotopes and the 1275 keV line from $^{22}$Na \citep{GomezGomar1998}. The closest nova in the last two decades, V5668 Sgr at 1--2 kpc from the Earth, is constrained by the observation by INTEGRAL/SPI for its $^{7}$Be mass $\lesssim 1.2 \times 10^{-8}~M_{\odot}$ \citep{Siegert2018, Siegert2020}. A similar approach using CGRO/COMPTEL is done for Nova Cygni 1992 which is constrained to have a $^{22}$Na mass $< 2.1 \times 10^{-8}~M_{\odot}$.

SN 2014J is the only SN Ia 
directly observable in $\gamma$-ray. $\gamma$-ray photons are detected $\sim10$ days after explosion and the observed line velocity of $^{56}$Co \citep{Diehl2015} has indicated the explosion asymmetry \citep{Leung2021SN2014J} (but also see \citet{Churazov2015} for other interpretations).

The well observed SN 1987A is the only CCSN with documented $\gamma$-ray signals. However, the resolution at that time does not show a clear line profile at early time \citep[see e.g., ][]{Sunyaev1990, Sunyaev1991, Pinto1988}, until last decade \citep{Boggs2015, Grebenev2012}. The supernova remnant Cassiopeia-A is the only remnant detected with a significant flux of $^{44}$Ti \citep{Iyudin1994, Tsygankov2016, Weinberger2020}, with decay lines at 68, 78 and 1157 keV. With the expected explosion happened around 1681 $\pm$ 19 years \citep{Fesen2006} the strong flux indicates a high $^{44}$Ti mass $\sim 1$--$2 \times 10^{-4}~M_{\odot}$ \citep{Siegert2015, Weinberger2020}. 

The possibilities of detecting $\gamma$-ray signals are optimistic thanks to the future proposals including e-ASTROGRAM \citep[0.3--3 GeV;][]{DeAngelis2017}, AMEGO \citep[All-Sky Medium Energy Gamma-Ray Observatory: 200 keV--20 GeV;][]{Kierans2020}, LOS \citep[Lunar Occultation eXplorer: 0.1--10 MeV;][]{Miller2019}, and the recently accepted COSI mission \citep[Compton Spectrometer and Imager: 0.2--5 MeV;][]{Tomsick2019}. They will search the $\gamma$-ray lines emitted from radioactive isotopes \citep[see ][for an overview of future $\gamma$-ray telescope projects]{Isern2021}. The multiple projects with different sensitivities and bandwidths will offer valuable chances to discover both direct and diffused $\gamma$-ray lines from the potential supernova candidates. 
In particular, the proposed $\gamma$-ray telescope COSI, which will be launched in 2025, made use of sixteen high-resolution germanium detectors. It can perform imaging of the sky in soft $\gamma$-ray surveys \citep{Zoglauer2021}. Its two-year mission can provide sensitivities by a factor of 10 or above compared to that of SPI and COMPTEL \citep{Siegert2022}. The high resolution will enable us to identify much more $\gamma$-ray sources. This motivates us to model the potential $\gamma$-ray signals from these sources systematically.


We envision the code to possess their features: (1) Being lightweight for doing parameter surveys in a reasonable computational time; (2) being portable to process models from different sources; (3) being flexible to include different input physics and parameters without extensively restructure the code. We choose to code in Python because Python has a broad user base, rich supporting libraries and flexible data structure suitable for mapping different types of source data\footnote{To the author's understanding there is no Python code designed for the modeling of $\gamma$-ray spectra}.  

In this article, we first present the algorithm and the code structure in Section \ref{sec:methods}. From Sections \ref{sec:num_comp} to \ref{sec:ccsn} we apply the code in multiple scenarios, including code tests for essential components, the post-explosion $\gamma$-ray emission in nova, SN Ia and CCSN explosions. We also compare our results with some representative $\gamma$-ray radiative transfer results. In Section \ref{sec:discussion} we explore the sensitivity of our code with different input physics and formulae applied for microphysics. Finally we give our conclusion.

\begin{table} 
    \centering
    \caption{The nuclear networks used for nucleosynthesis in this work. The column corresponds to the project. The isotopes which are the major radioactive decay sources are listed. }
    \begin{tabular}{c c c c c c}
    \hline
        element & Z & Nova & SN Ia & CCSN & $\gamma$-source? \\ 
        section & & \ref{sec:co_novae}, \ref{sec:one_novae} & \ref{sec:snia} & \ref{sec:ccsn} & \\ \hline
        hydrogen & 1 & 1-3 & 1-3 & 1-2 & \\
        helium & 2 & 3-4 & 3-4 & 3-4 & \\
        lithium & 3 & 7 & 6-7 & 6-7 & \\
        beryllium & 4 & 7, 9-10 & 7-9 & 7, 9 & 7 \\
        boron & 5 & 8 & 8-11 & 8, 10-11 & \\
        carbon & 6 & 12-13 & 11-14 & 11-13 & \\
        nitrogen & 7 & 13-15 & 12-15 & 13-15 & 13 \\
        oxygen & 8 & 14-18 & 14-19 & 14-18 & 15 \\
        fluorine & 9 & 17-19 & 17-21 & 17-19 & 18 \\
        neon & 10 & 18-22 & 17-24 & 18-22 & \\
        sodium & 11 & 21-24 & 19-27 & 21-23 & 22 \\
        magnesium & 12 & 23-26 & 20-29 & 22-27 & \\
        aluminium & 13 & 25-27 & 22-31 & 25-29 & 26 \\
        silicon & 14 & 27-28 & 23-34 & 26-32 & \\
        phosphorous & 15 & 30-31 & 27-38 & 27-34 & \\
        sulfur & 16 & 31-32 & 28-42 & 30-37 & \\ \hline
        chlorine & 17 & -- & 31-45 & 32-38 & \\
        argon & 18 & -- & 32-46 & 34-43 & \\
        potassium & 19 & -- & 35-49 & 36-45 & 40 \\
        calcium & 20 & -- & 36-49 & 38-48 & \\
        scandium & 21 & -- & 40-51 & 40-49 & 44 \\
        titanium & 22 & -- & 41-53 & 42-51 & 44 \\
        vanadium & 23 & -- & 43-55 & 44-53 & 48 \\
        chromium & 24 & -- & 44-59 & 46-55 & 48 \\
        manganese & 25 & -- & 46-61 & 48-57 & 54 \\
        iron & 26 & -- & 47-66 & 50-61 & \\
        cobalt & 27 & -- & 50-67 & 51-62 & 56 \\
        nickel & 28 & -- & 51-68 & 54-66 & 56 \\
        copper & 29 & -- & 55-69 & 56-68 & \\
        zinc & 30 & -- & 57-72 & 59-71 & 65 \\
        gallium & 31 & -- & 59-75 & 61-73 & \\
        germanium & 32 & -- & 62-78 & 63-75 & \\ 
        arsenic & 33 & -- & 65-79 & 65-76 & \\
        selenium & 34 & -- & 67-83 & 67-78 & \\
        bromine & 35 & -- & 68-83 & 69-79 & \\ \hline
    \end{tabular}
    
    \label{table:xiso}
\end{table}

\section{Methods}\label{sec:methods}
\subsection{Microscopic Data}\label{sec:stellar_evolution}

\begin{table} 
    \centering
    \caption{Essential isotopes and their radioactive decay channels considered in this work. $Q$ is the $Q$-value of the decay \citep{TabRad_v8}.
    }
    \begin{tabular}{c|c|c|c|c}
        \hline
        isotope & half-life & $Q$ (keV) & channel & $\gamma$-rays (keV) \\ \hline
        $^{7}$Be & 53.12\,d & 477.6 & EC & $478$ \\
        $^{13}$N & 9.97\,min & 1200 & $\beta^+$ & $\leq 511$; 1198  \\
        $^{15}$O & 2.04\,min & 1735 & $\beta^+$ & $\leq 511$ \\
        $^{18}$F & 109.7\,min & 633.5 & $\beta^+$ & $\leq 511$; $634$ \\
        $^{22}$Na & 2.6\,yr & 1275 & $\beta^+$ & $\leq 511$; $1275$ \\
        $^{26}$Al & 7.15\,Myr & 1809 & $\beta^+$ & $\leq 511$; $1809$ \\ \hline
        $^{44}$Ti & 60.0\,yr & 267.5 & $\beta^+$ & $\leq 511$; $67.9$; $78.4$ \\
        $^{48}$V & 15.97\,d & 4012.3 & $\beta^+$ & $\leq 511$; $983$; $1312$ \\
        $^{48}$Cr & 21.56\,hr & 1659.8 & $\beta^+$ & $\leq 511$; $112.4$ \\ \hline
        $^{56}$Co & 77.2\,d & 4566 & $\beta^+$ & $\leq 511$; $847$; \\
        $^{56}$Ni & 6.10\,d & 2135 & EC & $158$; $812$ \\
        \hline
    \end{tabular}
    
    \label{tab:isotopes}
\end{table}

\begin{figure}
    \centering
    \includegraphics[width=8cm]{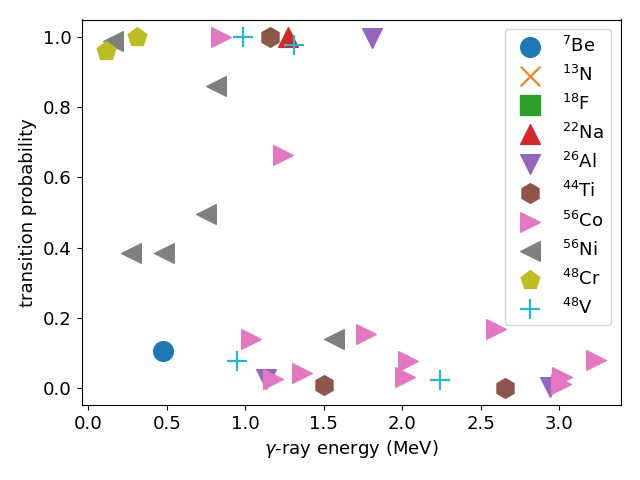}
    \caption{The transition probability against spectral line energy for the main radioactive isotopes included in the code.}
    \label{fig:spectral_line}
\end{figure}

In Figure \ref{fig:spectral_line} we list the  transition lines and probabilities of major radioactive isotopes used in the code for building the spectra\footnote{Nuclear data obtained online from \href{http://www.lnhb.fr/nuclear-data/nuclear-data-table/}{Laboratoire National Henri Becquerel}. Also refer to \cite{TabRad_v8} for the PDF version.}. Most elements have at least one strong line (transition probability $\approx 1$) from $\sim$100 to 2000 keV, with the highest energy released by $^{26}$Al at 1809 keV. Iron-group elements such as $^{56}$Co and $^{56}$Ni have much more weak lines where the transition probability is around 1--10\%, with $^{56}$Co having the largest number of transition lines. 

\subsection{Monte-Carlo Radiative Transfer}\label{sec:MC_radiative}

\begin{figure*}
    \centering
    \includegraphics[width=19cm]{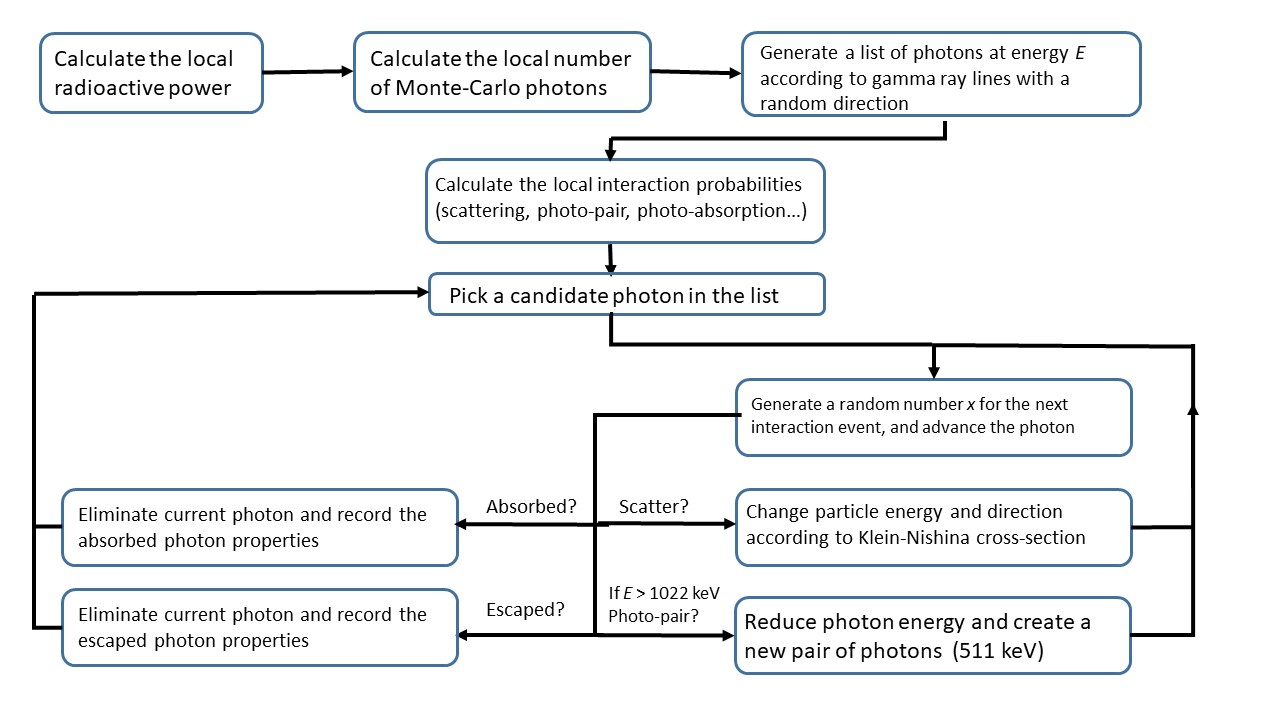}
    \caption{The preparation and iteration prescription of our $\gamma$-ray Monte-Carlo radiative transfer code. }
    \label{fig:code_flow}
\end{figure*}

Here we present the structure and the selected input physics used in our Monte-Carlo radiative transfer code. The qualitative design of the code is shown as a flowchart in Figure \ref{fig:code_flow}.
The code models how the radioactive nuclei generate $\gamma$-ray photon packets by radioactive decay, and how these packets experience the propagation, scattering and interaction with electron and nuclei in the ejecta.
In general, one needs to model all the $\gamma$-ray photons generated throughout the stars. However, at early time, most of the ejecta remains optically thick to $\gamma$-rays. The photons from the optically thick region are mostly scattered and absorbed. Therefore, in each time step, we search the optically thin region by estimating the optical depth of the matter by a grey opacity $\kappa_{\gamma} = 0.06 Y_e$ \citep{Swartz1995}. 
%
The mass shell with an optical depth $\tau = \int_r^R \rho(r') \kappa_{\gamma} dr' = 5$ is searched. Photons coming below that radius are assumed to be all scattered or absorbed. Only photons emitted above that is modeled. 
%
In the grey opacity limit, $\sim$99\% of the outward propagating photons from the $\tau=5$ surface are expected to be absorbed. 
%

The photon packet is generated by the following prescription. In each mass shell, we calculate the radioactive power per mass $q_{\rm decay}$ by adding the decay of all radioactive isotopes listed in Table \ref{tab:isotopes}.
The total radioactive power is obtained by summing all mass shells $\Delta m$ by $L_{\rm total} = \sum q_{\rm decay} \Delta m$ from the surface to the chosen innermost cell $\tau_{\gamma} = 5$ to compute the instantaneous $\gamma$-ray emission. 
Each photon packet corresponds to a collection of photons undergoing the same process during their propagation inside the star \citep{Ambwani1988}.
We treat each photon packet to be the fundamental unit in energy transport which cannot be further divided. \citep{Lucy2005}. When they scatter they only lose energy without splitting into multiple photons. The only exception is when the photon undergoes photopair production.
Each packet by definition corresponds to the `instantaneous photons emitted per unit time' instead of a physical photon.
Thus, one photon packet represents a luminosity $L_{\rm packet} = L_{\rm total} / N_{\rm packet}$. Here we assume that each packet has the same weight. This approach is appropriate for capturing the strong lines from radioactive nuclei. If the weak lines are of interest, an adaptive size of photon packets is necessary. 
For the scenarios in this work, we find that the time delay from photon emission to its escape is short so that the static approximation is appropriate. For more transparent or spatially extended ejecta, the propagation time becomes comparable with the interval between consecutive spectra snapshots. In that case, we need to account for the actual time-delay.

\subsubsection{Photon Generation}\label{sec:photon_generation}
In the frequency range considered, the $\gamma$-ray photons are assumed to be generated solely by the decay of radioactive isotopes\footnote{The $\gamma$-rays from excited nuclei will be an interesting but extensive feature to be added in the future.}. From given time snapshots of the ejecta, which include the kinematics, thermodynamics and isotopic distribution obtained from supernova or nova models, we calculate the spectra associated with these snapshots.
In Table \ref{tab:isotopes} we tabulate the principle parameters for the isotopes of interest.

If an isotope decays and then emits a photon, the code generates a photon packet by assigning it the corresponding $\gamma$-ray line energy with an arbitrary direction in the co-moving frame of the nuclei. The number and energy of photons are selected according to the available lines presented in Table \ref{fig:spectral_line}. The table can be easily extended by including the relevant parameters (half life, decay channels and probabilities, and the associated $\gamma$-ray emissions).
The $\gamma$-ray energies and directions are then transformed back to the lab frame.
The thermal fluctuation and collective motion of the nuclei in our calculations are small enough that the relativistic Doppler effects are small. 

Some isotopes are capable of decaying through $\beta^+$-decay. The Coulomb interaction from neighbouring electrons can make the positrons lose its energy quickly. It then captures an electron to form Positronium (Ps). 
%
%
%
Ps has two spin states: para-Ps and ortho-Ps.
The former emits two photons and the latter three due to charge and spin conservation \citep[see][for fundamental features of Ps]{Ore1949,Berko1980}.
Their relative ratio is limited by quantum statistics to a maximum of para-Ps:ortho-Ps$=$1:4.5.
The exact ratio depends on the thermodynamics of the electron including the matter density and temperature \citep[e.g.,][]{Leising1987}.
In our case, where the matter is opaque and dense, we expect the ratio to approach the quantum limit.
In the case of para-Ps (two-photon emission), we assign two photon packets of energy 511\,keV.
The first one has an arbitrary direction, with the direction of the second packet chosen by momentum conservation.
In the case of ortho-Ps (three-photon emission), the individual photon energy is chosen by the Monte-Carlo prescription descibed in \citet{Ore1949}, which is given by
\begin{eqnarray}
F = 2 \biggl[ \frac{x(1-x)}{(2-x)^2} - \frac{2 (1-x)^2}{(2-x)^2} \log(1-x) + \nonumber \\  \frac{2-x}{x} + \frac{2 (1-x)}{x^2} \log(1-x) \biggl],
\label{eq:Ore}
\end{eqnarray}
with $x = E_{\nu}/(m_ec^2)$ being the ratio of photon energy to the electron rest-mass energy.

In Appendix \ref{sec:prob_cdf} we describe how we construct the random number generator (RNG) which reproduces the given distribution. Other photon energies and directions are chosen by energy conservation (a sum of \,1022 keV) and momentum conservation (zero momentum). The numerical test of the random number generator is presented in next sections.
%

Once the energy of the three photons are determined, the equation set is closed and we can obtain the directions of the other two photon packets by momentum conservation.
The time-delay from the formation of Ps to its decay is $10^{-9}$--$10^{-6}$\,s \citep{Czarnecki1999}, which is much shorter than its escape time and the dynamical timescale of the ejecta. As remarked in \cite{Milne2004}, direct simulations of positron transport done in \cite{Milne1999} show that the escape of positrons from the ejecta before $\sim 150$ days is insignificant. The thermalization of positrons and their later annihilation occur, to a good approximation, on site and almost instantaneously. In this work, we modeled the $\gamma$-ray spectra up to 100 days after explosion. Thus, positrons are absorbed locally and the 2- and 3-photon generation is instantaneous in the code.


\subsubsection{Photon Interactions}\label{sec:photon_interactions}
We consider the following three types of interaction processes that change the energy of photon packets \citep{Pozdnyakov1983}: 

\noindent (1) \textbf{Compton scattering}: the photon packet transfers energy to electrons and lose energy.
The energy before scattering $E$ and after scattering $E'$ is related by \citep{Rybicki1985}:
\begin{equation}
    E' = \frac{E}{1 + (E/m_e c^2) \cos \theta}\mrm{,}
    \label{eq:compton_scattering}
\end{equation}
where $m_e$ is the electron mass and $\theta$ is the scattered angle in the center-of-mass frame. 
To calculate the post-scatter angle by a Monte-Carlo process, we use the Klein-Nishina formula which describes the differential cross-section of the relativistic Compton scattering, ignoring possible polarization, between $E$ and $E'$ by
\begin{equation}
    \frac{d\sigma}{d\Omega} = \frac{r^2_0}{2} E'^2 \biggl( \frac{E}{E'} + \frac{E'}{E} - \sin^2 (\theta) \biggl).
\end{equation}
The $r_0$ is the classical electron radius. The cumulative distribution function of the scattering $\sigma_{\rm CDF} = \int_0^{\theta_f} d\theta \int_0^{2\pi} d\phi (d\sigma/d\Omega)$ from $\theta=0$ to $\theta_f$ is given by the formula:
\begin{eqnarray}
    \sigma_{\rm CDF} = \frac{r^2_0}{4 x^2} \biggl[ 2 \theta - \frac{2 N_1(x) \arctan(\sqrt{1+2x} \tan(\theta/2))}{(1+2x)^{5/2}}  + \nonumber \\
    \frac{x^3 \sin \theta}{(1+2x)(1+x(1-\cos \theta))^2} + \frac{x N_2(x) \sin \theta}{(1+2x)^2 (1 + x(1 - \cos \theta))} \biggl],
    \label{eq:Ada}
\end{eqnarray}
with $x = E_i / m_e c^2$ being again the initial energy scaled by electron rest-mass, $N_1(x) = 11 x^4 + 4 x^3 - 12 x^2 - 10x - 2$ and $N_2(x) = 3 x^3 + 11 x^2 + 8 x + 2$ being the auxiliary functions \citep{Ada2014}.
Notice that when $\theta_f = \pi$ it represents the total scattering cross section for a given energy
\begin{eqnarray}
    \sigma_{\rm KN} = \frac{2 \pi r^2_0}{x} \biggl[ \frac{1+x}{x^2} \biggl( \frac{2 x(1+x)}{1 + 2x} - \log(1+2x) \biggl) +  \nonumber \\
    \frac{\log(1+2x)}{2x} - \frac{1+3x}{(1 + 2x)^2} \biggl].
\end{eqnarray}
In this work, the electrons are assumed to be cold. This means that the thermal velocity in the comoving frame is neglected. This assumption is valid in our calculation because the scattering process is assumed only in the optically thin region, i.e. near the surface, when the matter has entered homologous expansion. Such matter has a low thermal energy to rest-mass energy $k_B T/m_e c^2 \ll 1$.

When a scattering event occurs, we use the RNG to decide the output angle $\theta_f$ and also the output energy $E_f$. The RNG uses the same approach described in Appendix \ref{sec:prob_cdf}, which reproduces the cumulative distribution function of Eq. (\ref{eq:Ada}).
We again show the performance of our RNG for this cumulative distribution function in the next section. 
\newline 

\noindent (2) \textbf{Photopair production}: the photon packet is assumed to lose all its energy to an electron and the electron later emits an $e^-$-$e^+$ pair similar to the two-photon emission described above.
The cross section of this process is calculated according to the photon energy \citep{Hubbell1969}:
    \begin{eqnarray}
        \sigma &=& A (E - 1.022) Z^2 \times 10^{-27} {\rm cm^2}, ~ 1.022 < E < 1.5 {\rm MeV}\mrm{;} \nonumber \\ 
        \sigma &=& [B_1 + B_2 (E - 1.5)] Z^2 \times 10^{-27} {\rm cm^2}, ~ E \geq 1.5 {\rm MeV}\mrm{,}
        \label{eq:photopair_production}
    \end{eqnarray}
with $A$, $B_1$ and $B_2$ being 1.0063, 0.481 and 0.301 respectively.
$E$ is the photon packet energy in units of MeV. \newline

\begin{figure}
    \centering
    \includegraphics[width=8cm]{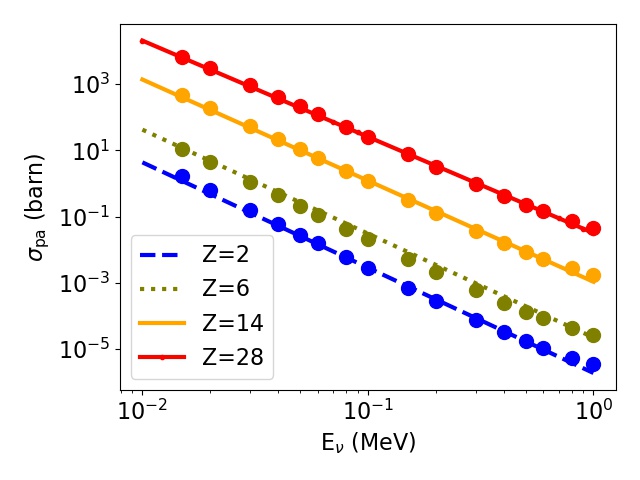}
    \caption{The numerical fittings of the photoelectric absorption cross sections for $Z = 2, 6, 12, 28$ and their comparison with observational data as circles.
    Numerical cross-section data are taken from \href{https://physics.nist.gov/}{NIST}.}
    \label{fig:fit_pe}
\end{figure}

\noindent (3) \textbf{Photoelectric absorption}: the photon packet is assumed to be absorbed by an electron.
The cross section takes the form $\sigma = \exp(C) E_{\rm MeV}^d$ with $C$ and $d$ being the parameters fitted from experimental data:
\begin{eqnarray}
    C &=& -0.011029Z^2 + 0.70509Z - 14.53767, \\
    d &=& 0.010592 Z - 3.20063.
    \label{eq:photoeffect}
\end{eqnarray}
The fitting is valid for $Z = 1$ to 30, applicable for energy from 0.01 -- 1 MeV. This range is sufficient for our purpose in the code as we expect $\gamma$-ray photons with an energy $< 100$ keV are mostly absorbed, while above $\sim$0.1--1 MeV the interaction is dominated by Compton scattering. 
Although the exact values of $C$ and $d$ depend on the fluorescence K-line, which is sensitive to the element, $\sigma \sim E^{-3}$ holds true for a wide range of elements after the transition. This allows us to represent the structure of the cross section with this formula with a good accuracy. In Figure \ref{fig:fit_pe} we plot our numerical fitting compared with the experiment data for $Z=2,6,12,28$ for $E_{\gamma}$ from 0.01 to 1 MeV. The observational data are also included as circles. The two sets of data overlap in general with each other to a good accuracy. In the energy range we are interested, our formula provides a general fitting to most chemical elements we are concerned. 

The interaction of photons is also determined by random process. 
To determine when and which process takes place, in each step, we assign a random number $x \in (0,1)$ so that  $\Delta \tau = -\log x$ corresponds to the optical depth change until the photon encounter an interaction event. 
The distance traveled by the photon packet is estimated by 
the mean free path $\Delta r_i = \Delta \tau / (\rho \kappa_i)$ for $\kappa_i$, where all three processes are considered.
By finding the minimum $\Delta r_i$, we assign the corresponding probability for each process to take place.
%
%
In the case where the traveling distance crosses the mass shell (defined by the stellar evolution model), we also update the local thermodynamical properties experienced by the photon packet. 

\subsection{Limitation of the Code}

This current version of code has a few major assumptions. Here we outline these assumptions and describe our reasoning and also the limitation by these assumptions.

The first assumption is that we assume the ejecta is spherically symmetric. We do so because the primary aim of the code is to model the $\gamma$-ray signature from the spherical symmetric nova models evolved from MESA as reported in \citep{Leung2021Nova}. The code aims at providing a flexible matching of that code, where the stellar evolutionary models with different setups can be mapped to our code easily. Thus, spherical symmetry is assumed in the first place. As remarked in \cite{Diehl2014}, some supernovae exhibit aspherical explosions. Extension is necessary to model these scenarios.

The second assumption assumes that the ejecta has developed homologous expansion profile for models we directly extrapolate in time. This is a good approximation for SNe Ia and novae where the ejecta mass is low. It typically takes less than days for the velocity profile to become time-independent \citep[see e.g.,][]{Roepke2007}. Simulations of supernova ejecta including radioactive decay as an energy source shows that the change of velocity due to this is secondary \citep{Blinnikov2006}. Assuming homologous expansion may eliminate the needs for evolving the fluid motion. When the ejecta is not fully in homologous expansion, direct simulations using radiation hydrodynamics are necessary to run the model to obtain the detailed density and velocity profiles for this code.

The last assumption is that the ejecta is cold, and all scattering channels are not sensitive to temperature. This is a good approximation for ejecta at days after explosion, where the matter becomes non-relativistic. However, the temperature dependence will be important for early time evolution and for scenarios like the gamma-ray burst. The inner core can remain $\geqslant 10^9$ K where thermal effects are important.  

\section{Numerical Components}
\label{sec:num_comp}
In this section we examine the numerical performance for some fundamental  components used in the code, including the distribution of the Klein-Nishina cross section, the Ps energy spectra and the  energy of $\gamma$-ray photons after single and multiple scattering. In Appendix \ref{sec:csball} we also present the spectra of a massive $^{137}$Cs-ball.

\subsection{Positronium Energy Spectrum}

We test the RNG for the Ps energy distribution. In \citet{Ore1949} the analytic formula is presented for the energy spectrum produced by Ps-annihilation. Notice that in the prescription, there are 6 unknowns, 3 for the energies of the 3 photons ($E_1, E_2, E_3)$ and the corresponding directions ($\theta_1, \theta_2, \theta_3$), while there are 3 constraints, 1 from energy conservation and 2 from momentum conservation -- assuming that all motion is confined on the $x$-$y$ plane. The choice of $\theta_1$ is arbitrary as it fixes the orientation of the system. The remaining 2 unknowns are determined by randomly choosing $E_1$ and $E_2$ which satisfy Eq. (\ref{eq:Ore}). Notice that in order to have a real solution for $\theta_i~(i=1,2,3)$, $E_i + E_j > 511$ keV $(i,j=1,2,3)$ is required. 

\begin{figure}
    \centering
    \includegraphics[width=8cm]{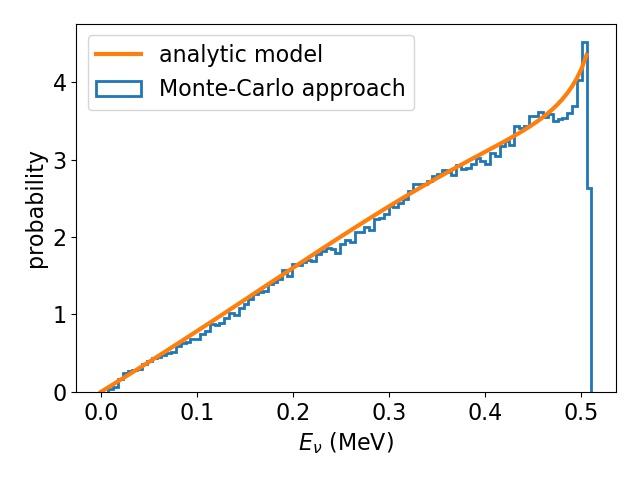}
    \caption{The numerical distribution of $N$ photon packets from decayed Ps and comparison with the analytic formula Eq.(\ref{eq:Ore}). Here we choose $N = 100000$.}
    \label{fig:spectra_ore}
\end{figure}

In Figure \ref{fig:spectra_ore} we compare the distribution function of our tuned RNG specific for the 3-photon scenario. We generate $3N$ photons and obtain their distribution according to energy. Most features for $E_{\gamma} = 0 - 0.46$ MeV are well captured by the RNG. Near $E_{\gamma} \approx 0.28$ and $0.50$ MeV, the RNG misses the analytic one by $\sim$10\%. We find that the rapid increase in the probability density function requires a higher order polynomial, or multiple functions to precisely describe that sharp rise. Despite that, the overall spectra are not sensitive to the detailed choice of the polynomial.

\subsection{Klein-Nishina Cross-Section}

The Compton scattering is the major channel for the high energy photons to cascade into lower energy ones with a continuum distribution. Here we test the RNG for the Compton scattering how it reproduces the Klein-Nishina cross section, where high energy photons tend to preserve its motion while lower energy photons are more likely to be reflected backward. 

\begin{figure}
    \centering
    \includegraphics[width=8cm]{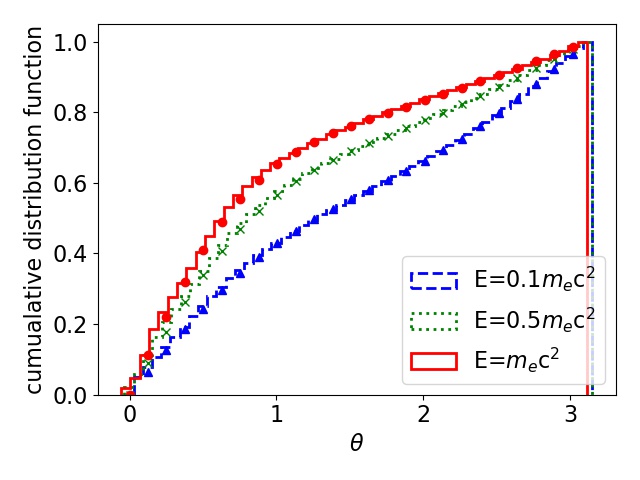}
    \caption{The numerical distribution of 100000 photon packets emerged from relativistic Compton scattering and the comparison with the analytic formula from Eq. (\ref{eq:Ada}). The data points are the expected values from analytic models.}
    \label{fig:spectra_ada}
\end{figure}

In the test, we use the RNG for the Compton scattering component to randomly predict the post-collision direction of $10^5$ photons for given energies. The cumulative distribution is plotted in Figure \ref{fig:spectra_ada}. The expected values from the analytic model (i.e., the cumulative distribution function) according to Eq. (\ref{eq:Ada}) are plotted as points on the figure. The overlap of the data points and the lines show that our RNG is consistent with the theoretical distribution. Moreover, the fast rising of the curve for higher energy particles agrees with the expectation that the photons tend to pass without significant energy transfer.

\subsection{Single and Multiple Compton Scattering}

Another test to understand the scattering component is to extract the energy distribution of the photons. The test aims at exploring the general properties of the energy under multiple scattering. 

\begin{figure}
    \centering
    \includegraphics[width=8cm]{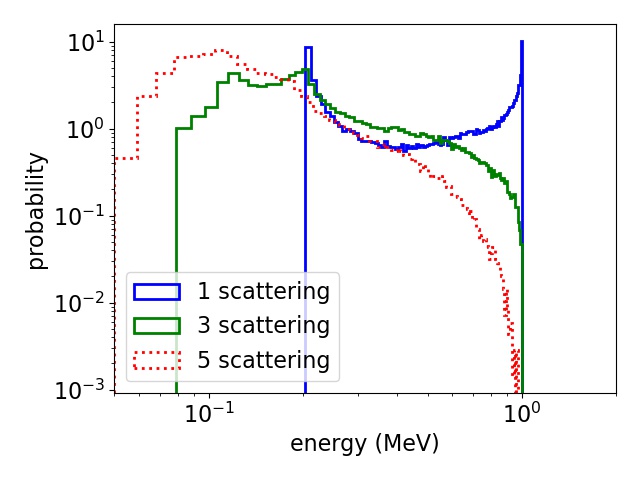}
    \caption{The energy distribution of $N$ 1-MeV electrons after 1, 3 and 5 scattering assuming static ejecta and no absorption. Here we choose $N = 100000$.}
    \label{fig:spectra_scatter}
\end{figure}

For each test, we prepare 100000 identical photons which have an initial energy of 1 MeV and random directions. They experience 1, 3 or 5 scatterings. The new direction and the new energy are determined by the RNG described above. The energy distribution of the forward moving photons are taken for forming the histogram shown in Figure \ref{fig:spectra_scatter}. The energy distribution of photons after 1 scattering agrees well with the Klein-Nishina distribution, where the peaks focus on both high and low energy -- photons are either moving forward as if there is no scattering, or it reverses by 180$^{\circ}$. 
When more than one scatterings occur, the high energy photons continue to cascade into lower ones. When the photons have experienced more scattering events, the energy spectra becomes more tilted at the lower energy side. A similar study in \citet{Brainerd1992}, albeit not in the exact configuration, also shows qualitatively similar features for photons experienced multiple Compton scattering.

\section{C+O Novae}
\label{sec:co_novae}

\subsection{Background and Method}

\begin{figure}
    \centering
    \includegraphics[width=8cm]{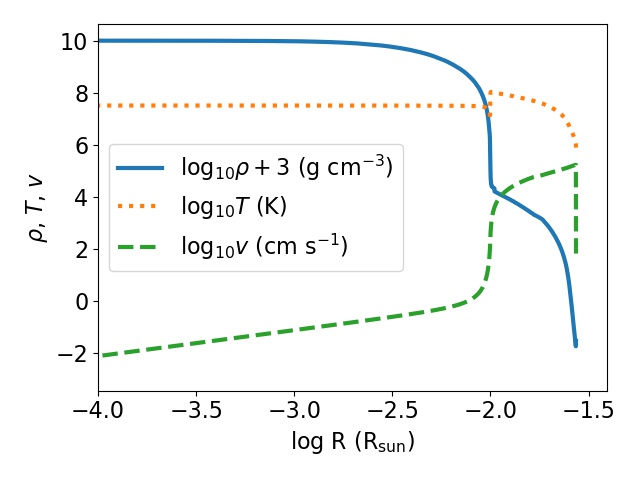}
    \includegraphics[width=8cm]{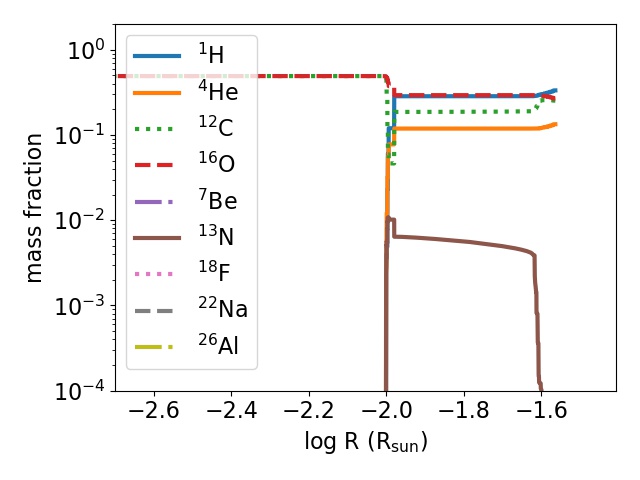}
    \caption{(top panel) The initial density, temperature and the velocity profiles of the CO080 model. (bottom panel) Same as the top panel but for the chemical abundance profile after the explosion. }
    \label{fig:CO080_init_model}
\end{figure}

Even though novae eject a small amount of mass during its outburst,
their occurrence rate is about 100--1000 times higher than supernovae in a galaxy; they are one of the robust sources for producing the diffusive $\gamma$-ray background in the galactic plane \citep{Diehl2021}.
Here we demonstrate the performance of the code using a representative C+O nova model. 
Similar to our previous works \citep{Leung2020cow,Leung2021SN2018gep,Leung2021Wave2}, we first prepare the stellar evolutionary model of a nova and then transfer the model for our calculation presented here. 
%
%
We choose the nova model when the mass outburst is occurring as the background model for the calculation of our $\gamma$-ray spectra.
%
%

To prepare the ejecta profile of a C+O WD, we use the stellar evolution code MESA (Module for the Experiments in Stellar Astrophysics) version 8118 \citep{Paxton2011, Paxton2013, Paxton2015, Paxton2018, Paxton2019}. The code solves the structure, nuclear reactions and radiative transfer inside a star with spherical symmetry.
The code first constructs a 3--7\,$M_{\odot}$ 
star and evolves till the formation of a CO WD
with a mass of 0.8\,$M_{\odot}$. We call this model CO080.
The WD is made to accrete C+O-rich matter until the thermonuclear runaway and outburst happen. 
We keep track of the synthesized radioactive isotopes of interest ($^{7}$Be, $^{13}$N, $^{14}$O, $^{15}$O, $^{18}$F, $^{22}$Na and $^{26}$Al) and the ejecta kinematics (see 
Table\,\ref{table:xiso}). We refer interested readers to \cite{Leung2021Nova} for the detailed prescription.

%
%
%
%
%

%

%
A moderate nuclear network (See Table \ref{tab:isotopes}) is used to keep track of the necessary radioactive isotopes (e.g., $^{18}$F, $^{22}$Na) while keeping the computational time feasible. In general low mass iron-group elements up to Ca is expected \citep{Jose2001}.
%

In Figure \ref{fig:CO080_init_model} we plot in the upper panel the initial hydrodynamical profile and in the lower panel the chemical abundance profile when the expansion starts. The C+O-rich core is not involved in the mass ejection process. It has a flat density profile for most part of the star and a steep density gradient near the interface. The temperature bump near $r \sim 10^{-2.0}~R_{\odot}$ corresponds to the position where the nuclear runaway takes place.\footnote{Notice that as the ejecta enters the homologous expansion, the ejecta temperature becomes irrelevant to the synthesis of the $\gamma$-ray spectrum. Thus, the code does not evolve the internal energy in the calculation.} Only the very outer part of the accreted matter has a high velocity to expand. 

The chemical isotope profile shows that  most of the accreted matter remains unchanged throughout the thermonuclear runaway. An extended layer of $^{13}$N by $^{12}$C$(p,\gamma)^{13}$N, spreads across the H-envelope through convective mixing. Most other radioactive isotopes are not seen because the host WD has a low mass, also low density at the interface. 

The first day of the mass ejection has a higher importance for novae because the decays of $^{13}$N and $^{18}$F are major sources of $\gamma$-ray photons. These isotopes have short half life times compared with the expansion time scale of the nova ($\sim 10$ days for our models).
These isotopes are distinctive from others because their main decay channel is $\beta^+$-decay. They emit positrons and (for the case of \nuc{O}{15}) energetic photons.

\subsection{Gamma-ray Radiative Transfer}

\begin{figure}
    \centering
    \includegraphics[width=8cm]{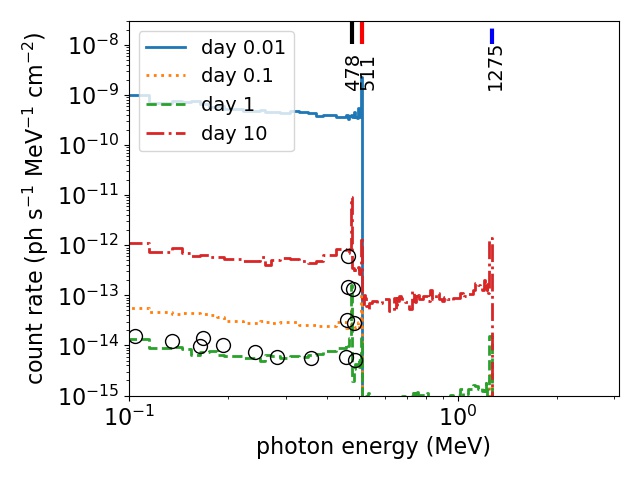}
    \includegraphics[width=8cm]{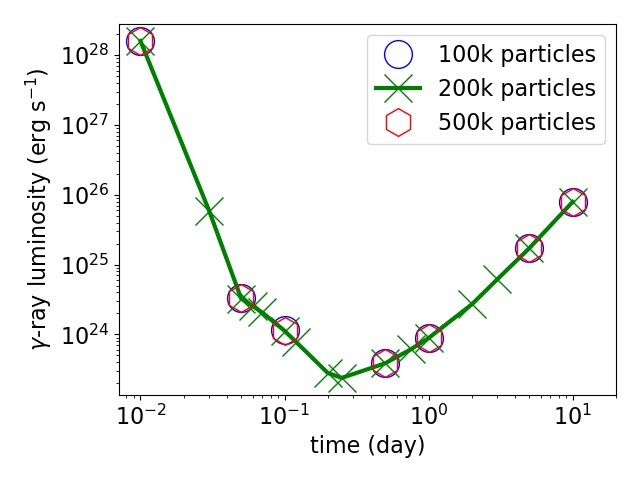}
    \caption{(top panel) The gamma ray spectra of the CO080 model at Day 0.01, 0.1, 1 and 10. The lines on the top corresponds to major decay energy lines. The circles correspond to the extracted spectrum from \citet{GomezGomar1998} for a similar C+O WD model scaled for comparison.
    (bottom panel) The gamma ray light curve of the CO080 model from Day 0.01 to Day 10 after the thermonuclear runaway. Models using other resolutions are also included for comparison.}
    \label{fig:CO080_gamma_plot}
\end{figure}

In the top panel of Figure \ref{fig:CO080_gamma_plot} we plot the $\gamma$-ray spectra of the ejecta. The global count rate decreases at early time due to the decay of very short lived isotopes such as $^{13}$N and $^{18}$F. The 511 keV line is also prominent. After that, the count rate sharply increases. The spectra show a rich background due to the multiple scattering of $\gamma$-ray photon in the opaque ejecta. The 487 keV line from $^7$Be decay becomes observable. There is a sharp drop in the count rate beyond 511 keV. Beyond Day 1, the spectra have a new line of 1275 keV from $^{22}$Na. The spectra shape remains barely changed between Day 1 and 10. 

In the bottom panel of Figure \ref{fig:CO080_gamma_plot} we plot the $\gamma$-ray light curve by integrating all the escaped photon packets. Consistent with the spectra, the light curve shows a sharp drop in at early time due to the decay of $^{13}$N, with a minor contribution of $^{18}$F. Then, as the photosphere slowly recedes, the luminosity gradually increases. The luminosity should reach its maximum after $\sim10$ days and converge to the total radioactive luminosity $L_{\rm decay}$. To demonstrate our results being insensitive to numerical settings, we repeat the numerical model but with different amounts of Monte-Carlo photon packets. The overlap of the data points show that the results are to a good approximation converged within the considered numbers of photon packets. 

Similar calculations of C+O nova $\gamma$-ray light curves and spectra have been made in \citet{GomezGomar1998}. With a different progenitor described in \citet{Jose1997}. The nova model experienced a stronger outburst when the WD has accreted more H-rich matter compared to ours, hence a much higher energy and expansion velocity. The main difference is the more updated radiative opacity adopted in the MESA code. Their ejecta becomes transparent at a much earlier time. Still, the sharp cut off at low energy, the later emergence of 478 keV line and 1275 keV line are common in both models. With the same reasoning, the evolution of the luminosity is much slower in our model that at Day 10 the light curve remains in the rising phase, while theirs has finished within 2 days. We compare our model around Day 1 with their model in Figure \ref{fig:CO080_gamma_plot} by scaling their spectra so that the strongest line aligns in its magnitude. The slope and the 478 keV line agree with each other.

\section{O+Ne Novae}
\label{sec:one_novae}

\subsection{Background and Method}

For stars with a mass 7--9 $M_{\odot}$, the stellar evolution ends at $^{12}$C-burning where a O+Ne rich white dwarf $\geq$ 1.0 $M_{\odot}$
is left behind as the remnant. The exact evolution of star in this mass range is less trivial because the O+Ne core can be degenerate, where the off-center burning of $^{20}$Ne and $^{28}$Si can bring additional chemical diversity in the WD \citep{Woosley2015}. Such burning might even trigger O+Ne flame (deflagration) which disrupts the WD by partial nuclear runaway or electron-capture induced gravitational collapse \citep{Jones2016, Zha2019, Leung2019ECSN}. The O+Ne WD in general leads to a stronger mass outburst \citep{Jose1998}.



The preparation of the outburst O+Ne WD model is similar to our C+O WD model but with a higher progenitor mass (1.2 $M_{\odot}$) and the accretion of O+Ne-enriched matter. The WD tends to outburst more frequently and strongly compared to the C+O WD models \citep{Leung2021Nova}.

\begin{figure}
    \centering
    \includegraphics[width=8cm]{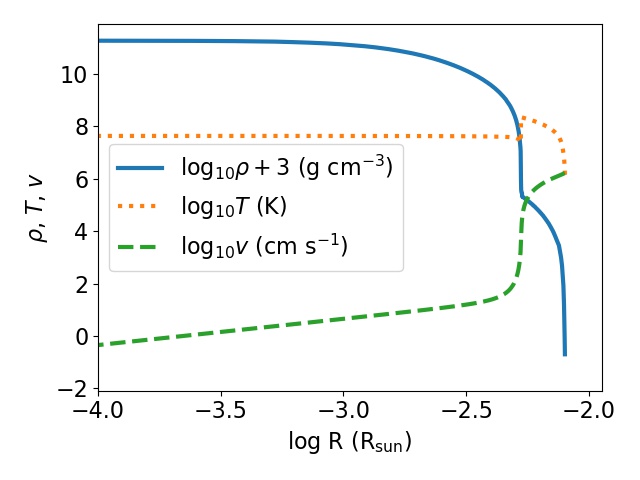}
    \includegraphics[width=8cm]{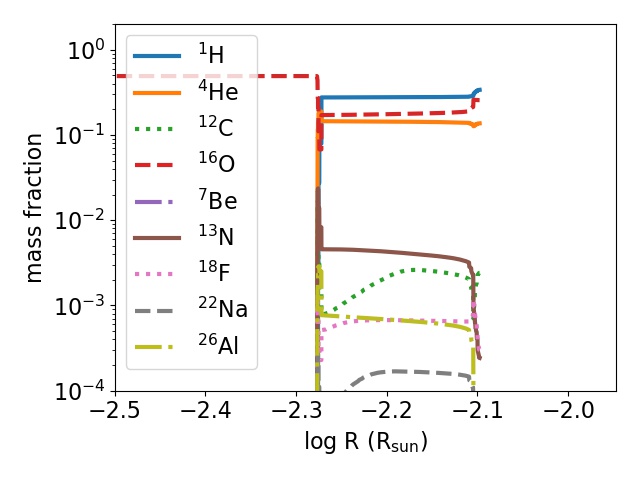}
    \caption{(top panel) The initial density, temperature profiles and the homologous expansion velocity profile of the ONe120 model. (bottom panel) Same as the top panel but for the chemical abundance profile after the explosion. }
    \label{fig:ONe120_init_model}
\end{figure}

In the top panel of Figure \ref{fig:ONe120_init_model} we show in the upper panel the hydrodynamics profile when the white dwarf begins to eject its H-envelope. The density and temperature structure are almost identical to the CO080 model in both the core and envelope. The temperature jump is marginally higher ($10^{8.3}$ K) and the entire envelope has a higher expansion velocity. 

In the bottom panel of the same figure, we show the chemical abundance profile of Model ONe120 at the same moment. The nucleosynthetic pattern shows more radioactive isotopes ($^{7}$Be, $^{18}$F and $^{22}$Na) compared to Model CO080. The isotopes are spread through the H-envelope by convective mixing. Such a mixing is important for early $\gamma$-ray signals because the ejecta expansion is slow and the decay half-lives of these isotopes are short. 

\subsection{$\gamma$-ray Radiative Transfer}

\begin{figure}
    \centering
    \includegraphics[width=8cm]{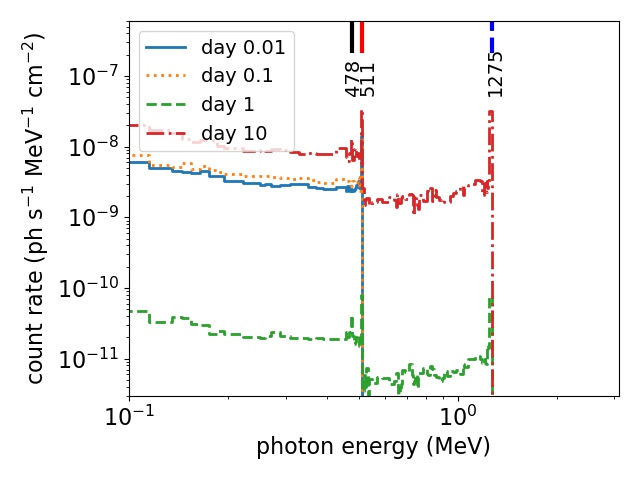}
    \includegraphics[width=8cm]{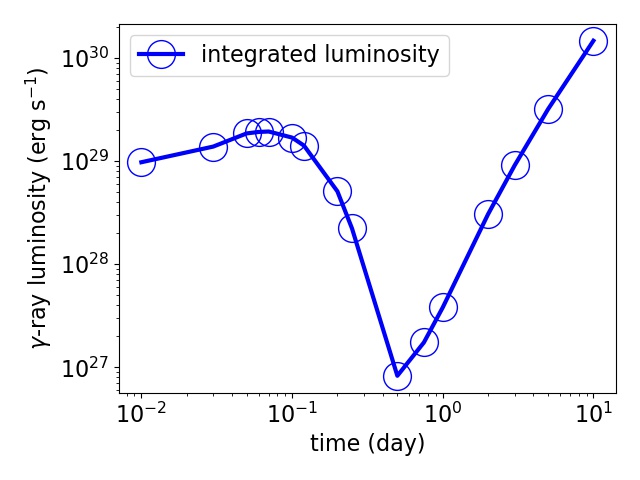}
    \caption{(top panel) The gamma ray spectra of the ONe120 model at Day 0.01, 0.1, 1 and 10. 
    (bottom panel) The gamma ray light curve of the ONe120 model from Day 0.01 to Day 10. We use the default $N=2 \times 10^5$ photon packets.}
    \label{fig:ONe120_gamma_plot}
\end{figure}

In the top panel of Figure \ref{fig:ONe120_gamma_plot} we plot the $\gamma$-ray spectra of the Model ONe120. The count rate for the same setting is larger for ONe120 than CO080. The early time features the very clear 511 keV line from$^{13}$N. It rapidly disappears after Day 1. The $^{7}$Be 478 keV line is strong in all the spectra. There is also a clear signal of the 1275 keV line from $^{22}$Na at Day 1 and beyond. The intensity of the spectra below 511 keV is slowly falling with photon energy. 

In the bottom panel of \ref{fig:ONe120_gamma_plot} we also show the integrated light curve at selected time. Different from Model CO080, the luminosity is flat in the first 0.1 day, then sharply drops to its minimum by almost three orders of magnitude around Day 0.5, and then gradually increases. The picture is consistent with the photosphere evolution as in the C+O nova described in the previous section.  

The O+Ne nova model is also studied in details in \citet{GomezGomar1998}. Using the stellar evolution code described in \citet{Jose1997}, which uses a stiffer radiative opacity, the O+Ne WD has a stronger outburst than our models. The higher explosion energy leads to a faster recession of photosphere. Still, some common features are consistent within the models: (1) the early 511 keV line and its later disappearance; (2) the later emergence of the $^{22}$Na 1275 keV line. 

\section{Type Ia Supernovae}
\label{sec:snia}

\subsection{Background and Method}

SNe Ia explode as the TNR of $^{12}$C and $^{16}$O, which primarily forms $^{56}$Ni. The high amount of $^{56}$Ni later decays into $^{56}$Co and $^{56}$Fe, which robustly emits $\gamma$-ray photons through electron capture ($^{56}$Ni, $^{56}$Co) and $\beta^+$-decay ($^{56}$Co). Small amount of $^{57}$Ni and $^{55}$Co are synthesized in the inner part of the star and some $^{48}$Cr in the outer ejecta. To present how the SN Ia model emits $\gamma$-ray photons, we consider the classical W7 model \citep{Nomoto1984}. The model considers the nuclear runaway of a near-Chandrasekhar mass WD $\sim 1.4~M_{\odot}$ where the burning is spread by a ``fast" turbulent flame, which scales with the local convective velocity. As discussed in the introduction, the observational data does not point at a unique explosion model. Despite limitations exist in the pure turbulent deflagration models, the W7 model has shown many interesting features which represent the general behaviour of typical SNe Ia, and has been used for $\gamma$-ray spectral synthesis in the literature \citep{Milne2004}. 

\begin{figure}
    \centering
    \includegraphics[width=8cm]{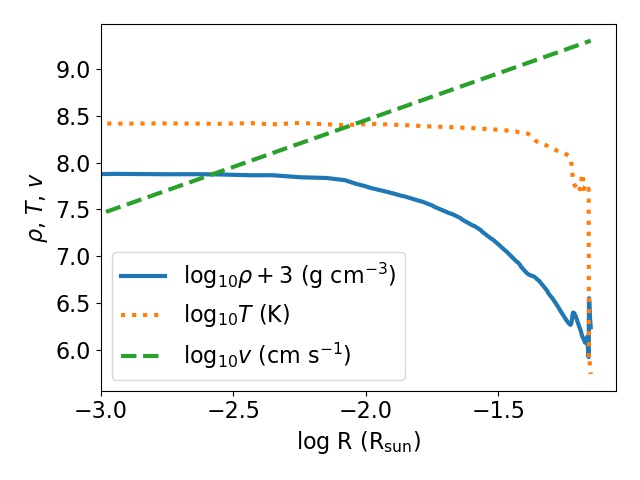}
    \includegraphics[width=8cm]{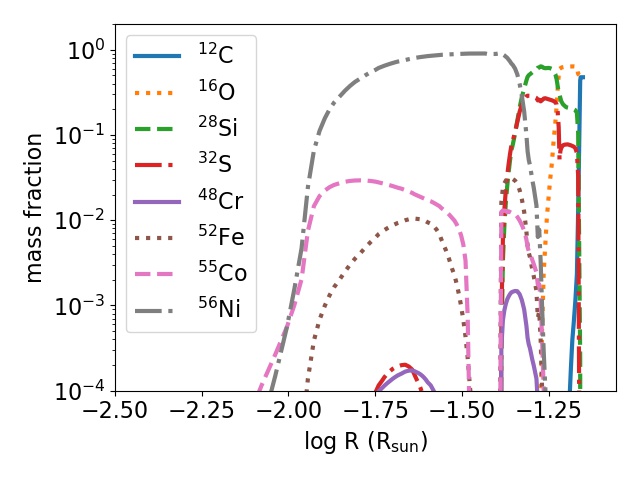}
    \caption{(top panel) The initial density, temperature and velocity profiles of the W7 model. (bottom panel) The chemical abundance profile after the explosion. }
    \label{fig:w7_init_model}
\end{figure}

The structure of the post-explosion WDs is still compact. The density and temperature are smooth due to the absence of supersonic detonation. There are small bumps in the density and temperature near the surface due to the sharp density gradient. The velocity near the surface reaches a few $\times 10^4$ km for the very outer part. In Figure \ref{fig:w7_init_model} we plot the initial hydrodynamical profile and the composition after the explosion for some representative isotopes. 

The chemical abundance pattern of the W7 model is taken from \cite{Nomoto2018} using the 495-isotope network with updated microphysics. It contains a rich amount of $^{56}$Ni which contributes to the majority of the ejecta. $^{55}$Co also contributes to the $\gamma$-ray source but with an abundance almost two orders of magnitude lower. Some radioactive $^{48}$Cr can be found at the outer $^{56}$Ni-layer. There is almost no low-mass radioactive isotopes such as $^{7}$Be and $^{13}$N because the initial composition is $^{12}$C and $^{16}$O-rich. 

\subsection{$\gamma$-ray Radiative Transfer}

\begin{figure}
    \centering
    \includegraphics[width=8cm]{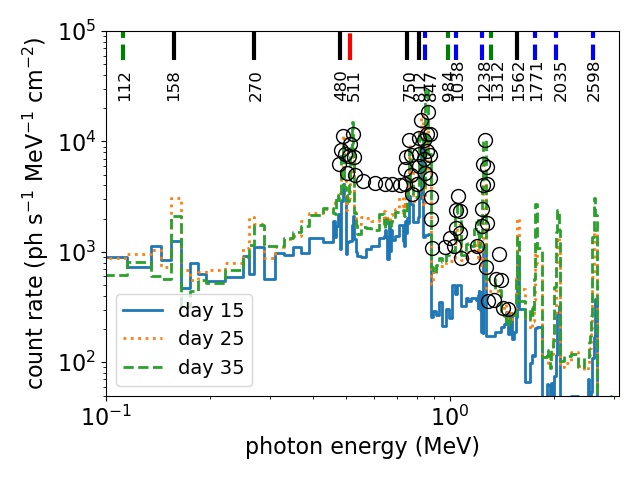}
    \includegraphics[width=8cm]{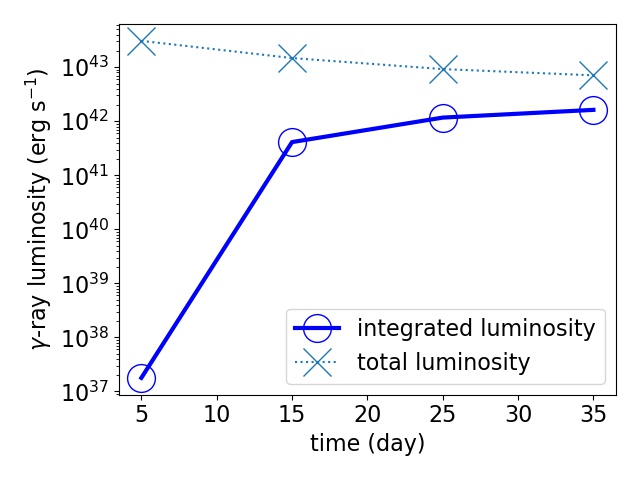}
    \caption{(top panel) The gamma ray spectra of the W7 model at Day 5, 15, 25 and 35. The circles are the data extracted from \citet{Milne2004} for the W7 model at a similar snapshot.
    (bottom panel) The gamma ray luminosity of the W7 model at Day 5, 15, 25 and 35. 
    }
    \label{fig:W7_gamma_plot}
\end{figure}

The code computes the spectra of any snapshot using the direct input files, such as the nova models presented above. The code can also extrapolate the profile in time when the ejecta assumes homologous expansion. In the top panel of Figure \ref{fig:W7_gamma_plot} we plot the $\gamma$-ray spectra of the W7 model at Day 5, 15, 25 and 35. No observed spectra at Day 5 as all the radioactive isotopes remained shielded inside optically thick layers. Prominent features include clear spectral lines at 158, 480, 750, 812 keV lines from $^{56}$Ni, and 511, 847, 1238, 2598 keV lines from $^{56}$Co. Double line features near 812 keV are well captured too. The drop of 811 keV line flux coming from $^{56}$Co is also observed.

In the bottom panel of Figure \ref{fig:W7_gamma_plot} we plot the $\gamma$-ray luminosity by integrating all escaped photons. The $\gamma$-ray luminosity increases sharply as the first light at around Day 15. It shows that the recession of the photosphere is significant to expose the radioactive isotopes. Notice that in a spherical explosion like this, (radioactive) Fe-group elements tend to be produced in the core. For comparison we also included the total luminosity defined by the total radioactive power of all isotopes inside the WD. The two lines rapidly approach, showing that the ejecta becomes close to transparent that the ejected photons equal to escaped photons. 

We compare our results with a similar SN Ia model described in \citet{Ambwani1988}. For their Chandrasekhar mass white dwarf model with $\sim 1.2$--$1.3~M_{\odot}$ ejecta with $0.6~M_{\odot}~^{56}$Ni, they observe strong lines in 511, 847 and 1238 keV, with the 1038 keV line being slightly weaker. The relative strength of these lines are consistent with theirs. The ``step-like'' pattern across the strong lines above and the power-law like decay of count rate beyond 1238 keV are also well reproduced. We also compare in the same figure the W7 model at Day 25 calculated in \cite{Milne2004} in the figure. Their spectra is scaled up accordingly by aligning the strongest line. Their results agree very much with ours including the line strength and width. Below 400 keV their scattering background  shows a higher count than our models. 
In Section \ref{sec:discussion} we show that the our generalized model for the scattering cross section for photoelectric absorption can reproduce the classical 3-element model described in \citet{Ambwani1988}. 

\section{Core-collapse Supernovae}
\label{sec:ccsn}

\subsection{Background and Method}

CCSNe are another important sources of $\gamma$-rays. The $\gamma$-ray spectra feature decay lines of $^{44}$Ti, $^{48}$Cr, $^{56}$Ni and $^{57}$Ni and their daughter nuclei. To demonstrate the code application in CCSN, we consider the N20 model. The model is evolved from a He star $\sim 6~M_{\odot}$ \citep{Shigeyama1988}. The model is proposed for explaining the optical features of SN 1987A. Compared to SN Ia models, the star has an extended He-envelope which extends to a few $R_{\sun}$. The original model has removed the H-envelope to match the blue progenitor of SN1987A. 
We use the recalculated model presented in \cite{Simionescu2019}.

\begin{figure}
    \centering
    \includegraphics[width=8cm]{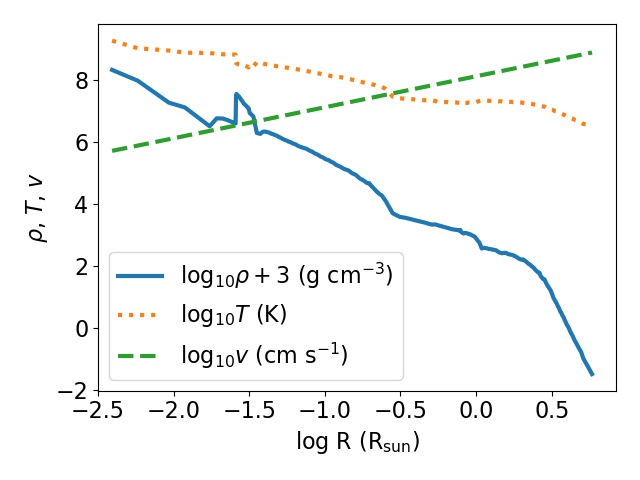}
    \includegraphics[width=8cm]{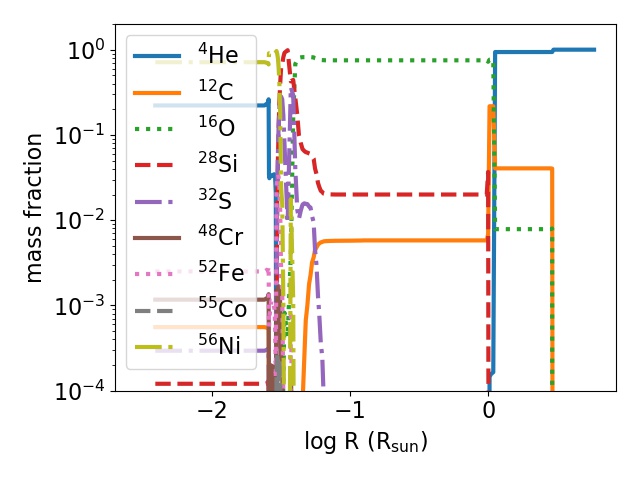}
    \caption{(top panel) The initial density, temperature and velocity profiles of the N20 model. (bottom panel) Same as the top panel but for the chemical abundance profile after the explosion. }
    \label{fig:N20_init_model}
\end{figure}

The explosion deposits an energy $\sim 10^{51}$ erg in the innermost 10 mass grids as a thermal bomb at the Si layer. It generates a shock and propagates outwards, triggering explosive nucleosynthesis. The density bump develops during the propagation of a shock wave across layers with different chemical elements. The He star has a more compact structure. Most radioactive isotopes, e.g., $^{56}$Ni ($\sim 0.07~M_{\odot}$) and $^{55}$Co ($1 \times 10^{-5}~M_{\odot}$) are in the innermost ejecta. Small fraction of $^{48}$Cr ($1 \times 10^{-4}~M_{\odot}$) and $^{44}$Ti ($4 \times 10^{-4}~M_{\odot}$) can be found in Si-rich layer. Almost no radioactive isotopes (e.g., $^{7}$Be, $^{13}$N) are found in the He layer as the shock has mostly dissipated. 

\subsection{$\gamma$-ray Radiative Transfer}

\begin{figure}
    \centering
    \includegraphics[width=8cm]{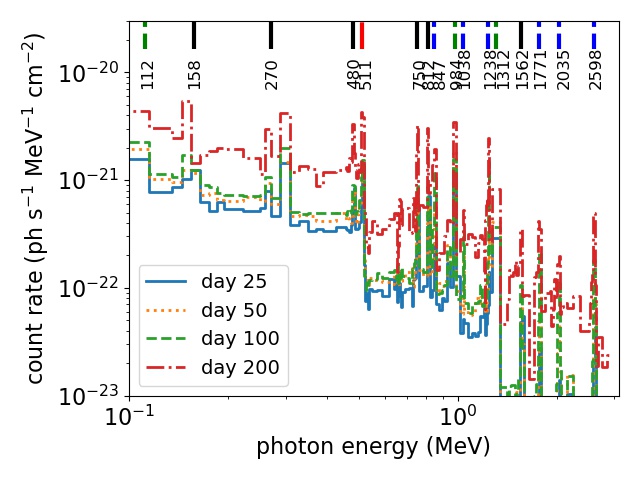}
    \includegraphics[width=8cm]{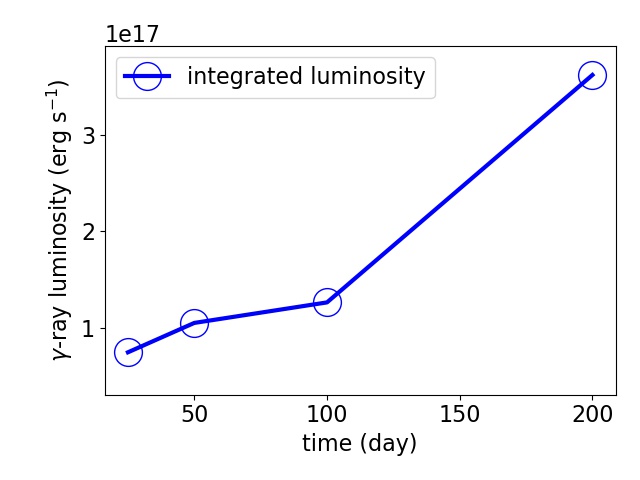}
    \caption{(top panel) The gamma ray spectra of the N20 model at Day 25, 50, 100 and 200.
    (bottom panel) The gamma ray spectra of the N20 model at Day 25, 50, 100 and 200. The particles are the number of the Monte-Carlo tracers of the $\gamma$-ray photon samples in each snapshot.}
    \label{fig:N20_gamma_plot}
\end{figure}

In the top panel of Figure \ref{fig:N20_gamma_plot} we plot the $\gamma$-ray spectra of the model at Day 25, 50, 100 and 200 respectively. The overall pattern does not change, with a peak around 100 keV and the intensity $I$ decreases with the photon energy ($I \sim E^{-2}$). The spectra contain contributions of multiple isotopes, including 112, 308 keV from $^{48}$Cr, 158, 480, 812 keV from $^{56}$Ni, 511 keV from e-capture of $^{48}$Cr and $^{48}$V, 847 and 1238 keV from $^{56}$Co and 1312 keV from $^{48}$V. The spectral luminosity is very low compared to the Model W7 counterpart. Similar to SNe Ia, most radioactive Fe-group isotopes are synthesized at the inner part of the ejecta. Unless the aspherical mixing is explicitly modeled, their radioactive power is heavily shielded by the envelope.  

In the bottom panel of Figure \ref{fig:N20_gamma_plot} we plot the total integrated luminosity of the escaped $\gamma$-ray photons at the same snapshot of the stellar profile.
The monotonic rising of the luminosity shows that the recession of the photosphere is faster than the decay half-life (100 day -- a few yrs) of the related isotopes. The absolute value of the $\gamma$-ray luminosity is much lower than novae and SNe Ia. This is expected due to the lower amount of radioactive isotopes synthesized outside the photosphere.
The low $\gamma$-ray luminosity for CCSN suggests that most photons cannot be observed within current telescope sensitivities beyond 50 kpc. However, when the ejecta enters the nebulae phase, the photons from very long-lived isotopes (e.g., $^{44}$Ti) can freely propagate around the medium. At that point, those photons become the diffused $\gamma$-ray photons with clear line features which can be observed.

\section{Discussion}
\label{sec:discussion}

\subsection{Sensitivity to Input Physics}

In \citet{Ambwani1988} the photoelectric absorption effect is modeled for three representative atomic number $Z = 7, 14, 28$, corresponding to unburnt fuel, partially incinerated ash and fully incinerated ash. Our work here provides a more generalized formula which extends from He to Zn based on the atomic cross section of individual elements. Here we examine how the fitting formula affects the spectral fit. We test by the W7 model. This model has a C+O rich surface and a Si-rich middle layer. But the atomic number does not exactly lie at the prescribed value. Therefore, we want to understand if the deviation due to transition across layers can cause deviation. 



\begin{figure}
    \centering
    \includegraphics[width=8cm]{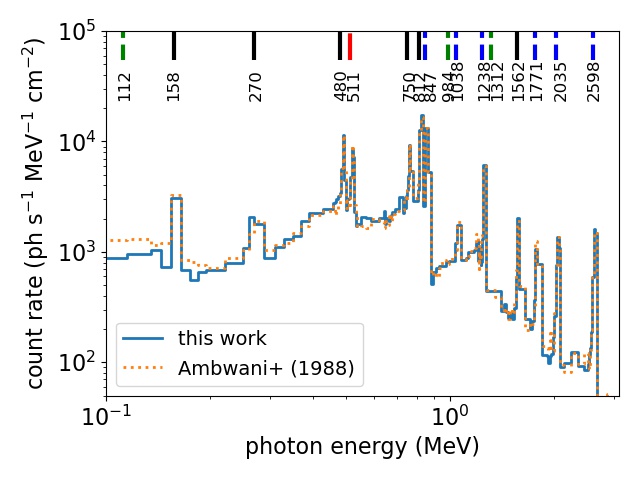}
    \caption{The spectra of the Model W7 at Day 25 computed using the fitting formula of photoelectric absorption in this work and by the formula presented in \citet{Ambwani1988}.}
    \label{fig:spectra_W7_comp_pe}
\end{figure}

In Figure \ref{fig:spectra_W7_comp_pe} we show the spectra at Day 25 using the two choices of fitting formula. The two models show a good agreement with each other except for minor deviations between 100 -- 200 keV. This confirms that our more generalized fitting can reduce to the three-element model described in \cite{Ambwani1988}. Minor difference can be observed near the lower end of the spectrum.

\subsection{Sensitivity to Numerical Setting}

We also test the approximations used in our code. An important one is the variable $\tau_{\rm max}$. We only take account the $\gamma$-ray photons emitted at layers above that layer of $\tau_{\rm max}$. Below that, their contribution is minute because these photons have to go through more scattering during their propagation towards the surface. The photons coming from these layers are likely to have sufficiently cascaded into lower energy photons, which are later absorbed by surrounding electrons through photoelectric absorption. It is expected that these photons, if escaped, contribute to the continuous spectra as in the optical band.

\begin{figure}
    \centering
    \includegraphics[width=8cm]{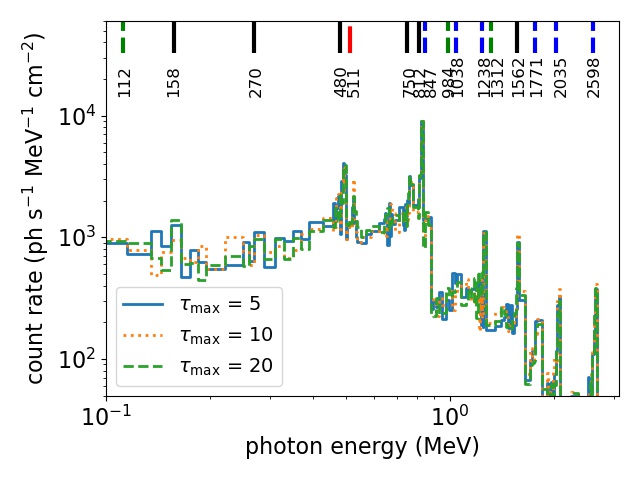}
    \caption{The spectra of the W7 model at Day 15 for $\tau = 5, 10, 20$ respectively.}
    \label{fig:spectra_W7_tau}
\end{figure}

In Figure \ref{fig:spectra_W7_tau} we show the spectra of the W7 model presented in Section \ref{sec:snia}. We repeat the calculation of the spectra at Day 15 but with different maximum optical depths. The three spectra agree with each other very well at all major lines at 480 keV onwards. Minor fluctuations can be seen but in general less than 10\% difference is found for photon energy $< 200$ keV.


The test confirms us that the current choice of $\tau_{\rm max} = 5$ is a sufficient choice to capture the important source of $\gamma$-ray in the ejecta. 

\subsection{Future Works}

This work focuses on the early time $\gamma$-ray signature. It is possible to extend the code to model the spectra at a later time. However, it requires a number of extensions: The code needs to take into account the propagation time of photons from their sources until they escape,  as well as the time-dependence of the ejecta during their expansion \citep[see the comparison in][]{Milne2004}. In this work, these effects remain small (e.g., less than a few seconds for novae and less than a day in SNe Ia) and the correction in the local thermodynamics is small. When the ejecta enters the nebula phase, photons travel for a significant amount of distance before the next interaction. The local thermodynamics condition experienced by the photons then depends on the arrival time of the photons at the shell considered. Besides the change of profile, to compute the instantaneous luminosity, the integration should include consistently the time delay. 

Complication occurs for microphysics in the nebula phase too. In the code, it is assumed that a positron is slowed down instantaneously by surrounding electrons by Coulomb interaction and forms Ps which also decays instantaneously. The assumption is valid at early time, where most matter remains mostly ionized. When the ejecta enters the nebula phase, similar to photons, the capture of electron for annihilation becomes non-local. To consistently model this phase, the positron itself should be modeled as another type of ``packet'' (while the annihilation of Ps is still much shorter than any dynamical timescales in a supernova). The extension will allow the code to model multiple possible interactions depending on the electron sources \citep[see e.g.,][for the possible interaction channels.]{Prantzos2011}.

This work assumes spherical symmetry in the ejecta distribution. For aspherical ejecta, the Monte Carlo approach can be naturally extended to multi-dimensional models, when the velocity profile is homologous expanding along each radial direction. The aspherical distribution of matter may allow mixing of radioactive elements (e.g., $^{48}$Cr, $^{56}$Ni) from the inner ejecta outwards. Their exposure may generate observable lines at early time. 
To extend the simulation dimension with the homologous expansion approximation, the code needs to store the stellar profiles in both radial and angular directions. The code needs to account for the migration of photon packets along the angular direction.

\subsection{Conclusion}\label{sec:conclusion}
We have presented a new Monte-Carlo radiative transfer code for $\gamma$-ray spectral line formation in Python. The code is designed with the principles being light-weight, portable and flexible. We have shown how the code can be applied in major $\gamma$-ray scenarios including C+O and O+Ne novae, Type Ia supernovae and core-collapse supernovae evolved from different codes. The code reproduces features from characteristic nova and supernova models reported in our previous works and other works from the literature.  

We have also done a number of code tests to validate the code. We demonstrate how the random number generator components can reproduce the analytic distributions in microphysics including the photon energy from Ps decay and the directional dependence of relativistic Compton scattering. We also studied how our results being insensitive to the choice of resolution and some numerical parameters. 

In the future the code will combine with our supernova modeling pipeline to generate $\gamma$-ray spectra, based on more diversified and systematic arrays of nova and supernova models obtained from stellar evolution and hydrodynamics simulations. The unified approach allows us to examine the effects of microphysics to various classes of transient objects. A set of these results will provide a consistent approach to predict how different types of supernovae generate the diffused $\gamma$-ray background in the galactic scale. These results will be important for the future $\gamma$-ray surveys and imaging projects, such as COSI.  

\section*{Acknowledgments}
S.C.L. acknowledges support by NASA grants HST-AR-15021.001-A and 80NSSC18K1017.
S.C.L. thank Thomas Siegert for the encouragement and many ideas during the development of this code. 

\section*{Data Availability}

The data underlying this article will be shared on reasonable request to the corresponding author. The source code is available on Zenodo
(\href{https://doi.org/10.5281/zenodo.6578600}{10.5281/zenodo.6578600}).

\noindent This project is done with the use of Python libraries: Matplotlib \citep{Matplotlib}, Pandas \citep{Pandas}, Numpy \citep{Numpy}, Scikit-Learn \citep{Sklearn}.


\bibliographystyle{mnras}
\pagestyle{plain}
\bibliography{biblio}

\appendix
\section{Appendix: Constructing the Random Number Generator}
\label{sec:prob_cdf}

In the main text the calculation of relativistic Compton scattering and the generation of photons by Ps-decay requires the use of a random number generator with a specific probability distribution function ($y = f_{\rm PDF}$, PDF). In our work here the PDF only depends on a single parameter, $x = E_{\nu} / m_e c^2$ being the scaled photon energy before collision for the Ps-annihilation and $\theta$ the photon angle for the Compton scattering. 
Here we outline the prescription used for converting a uniformly distributed random number generator $\tilde{N}$ to the one with a specified probability density function.

Assume a PDF $y = f_{\rm PDF}(z)$ is a function of a parameter $z$ valid for a domain in (0,$z_f$). We define the cumulative distribution function $F_{\rm CDF}$ as
\begin{equation}
    F_{\rm CDF}(z) = \int_0^{z} f_{\rm PDF}(z') dz'
\end{equation}
Then we invert the function $Y = F_{\rm CDF} (z)$ so that 
\begin{equation}
    z = F^{-1}_{\rm CDF} (Y),
\end{equation}
the required mapping function to convert $\tilde{N}$ into a random number with a distribution satisfying $f_{\rm PDF}$ is 
\begin{equation}
    z(\tilde{N}) = F^{-1}_{\rm CDF} (\tilde{N}).
\end{equation}    
    
When the inversion cannot be represented analytically, we use a power series $g(Y) = \sum^N_{i=0} a_i Y_i$ for some constants $a_i$. We find that a polynomial of order 7--10 is necessary to reproduce most of the observed features in the PDF.
    
\section{Numerical Test: A $^{137}$Cs ball}
\label{sec:csball}
There are two options in the code for setting up the initial models: (1) to read the predefined models evolved from other stellar evolution or supernova explosion codes and (2) to construct the envelope-like profile with a given energy $E$ and mass $M$. Here we demonstrate how the code solves option (2).

For a given envelope profile satisfying $\rho = \rho_0 / r^n$  valid for $(R_{\rm in}, R_{\rm out})$ with $0<n<2$ or $ n>5$, its mass is given by the formula
\begin{equation}
     M = \frac{4 \pi \rho_0}{3 - n} [R^{3-n}_{\rm out} - R^{3-n}_{\rm in}].
\end{equation}
Assume the envelope expands with a homologous expansion profile $v(r) = v_0 (r-R_{\rm in}) / (R_{\rm out} - R_{\rm in})$. The total kinetic energy of the ejecta is given by
\begin{equation}
    E = 2 \pi \rho_0 v_0^2 R_{\rm out}^{5-n} \biggl[ \frac{1}{5-n} - \frac{2 \beta}{4-n} + \frac{\beta^2}{3-n} - \Delta \beta^{5-n} \biggl],
\end{equation}
where $\beta = R_{\rm in}/R_{\rm out}$ and $\Delta = 1/(5-n) - 2/(4-n) + 1/(3-n)$. 
When $n$, $M$, $E$, $R_{\rm in}$, $R_{\rm out}$ and the composition are chosen, we solve for $\rho_0$ and $v_0$. Then we construct the density, temperature and chemical abundance profiles accordingly. 

\begin{figure}
    \centering
    \includegraphics[width=8cm]{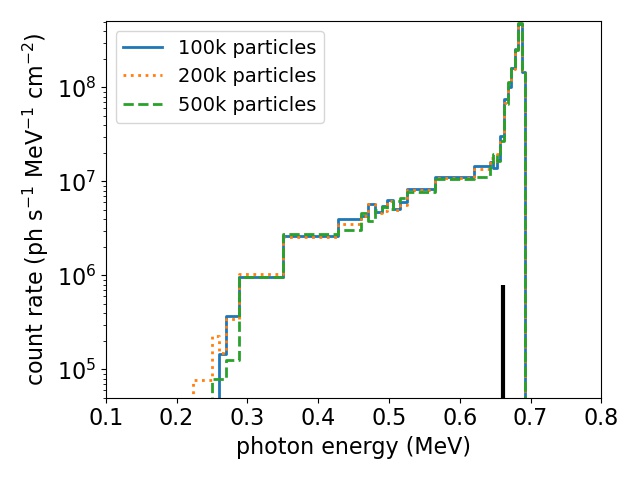}
    \caption{The energy spectrum of a fictitious $1~M_{\odot}~^{137}$Cs ball with a flat density profile and a total kinetic energy $10^{50}$ erg. Three resolutions, $10^5$, $2 \times 10^5$ and $5 \times 10^5$ photon packets are used for constructing the spectra. The black solid line corresponds to the 662 keV line.}
    \label{fig:spectra_CS137}
\end{figure}

As an example, we construct a $1~M_{\odot}$ star made of pure $^{137}$Cs expanding with a total energy of $10^{51}$ erg. $R_{\rm in}$ and $R_{\rm out}$ are chosen to be $10^{13}$ and $10^{15}$ cm. In Figure \ref{fig:spectra_CS137} we show the spectra at Day 0. $^{137}$Cs has only one strong line at 661.7 keV. 

The spectrum shows a very simple structure that a very sharp around the expected transition frequency. Then the scattering creates lower energy photons, where some sharp cutoff appears below $\sim 30$ keV by photoelectric absorption. 
Given the explosion energy and mass, the outermost matter is ejected with a velocity $\sim 0.04c$. This corresponds to a blueshift of the spectral line by $661.7 \sqrt{(1+\beta)/(1-\beta)} \approx 688.7$ keV. Notice that only blueshifted photons form the spectral line, the redshift one must experience one $180^{\circ}$ backward propagation before they can leave the ejecta. They form the low energy background in the spectrum.

We further test the code by repeating the calculation by different numbers of photon packets, from $10^5$ to $5 \times 10^5$. The three spectra almost overlap with each other by observation. Minor differences are found near the very low end of photon energy due to the statistical fluctuations.
    
\bsp	
\label{lastpage}
\end{document}